\documentclass[%
 reprint,
superscriptaddress,
nofootinbib,
 amsmath,amssymb,
 aps,
]{revtex4-2}

\usepackage{graphicx}
\usepackage{dcolumn}
\usepackage{bm}
\usepackage{amssymb}
\usepackage{amsmath}
\usepackage{amsfonts}
\usepackage{graphicx}
\usepackage{color}
\usepackage{xspace}
\usepackage[normalem]{ulem}
\usepackage{mathtools}
\usepackage{hhline}
\usepackage{braket}
\usepackage{subcaption}
\usepackage{float}
\usepackage[dvipsnames]{xcolor}
\usepackage{url}

\begin{document}


\title{Astrophysically sourced quantum coherent photonic signals}

        \author{Arjun Berera}
        \email{ab@ed.ac.uk}
        \affiliation{School of Physics and Astronomy, University of Edinburgh, Edinburgh, EH9 3FD, United Kingdom}
        
        \author{Jaime Calder\'on-Figueroa}
        \email{jaime.calderon@ed.ac.uk}
         \affiliation{School of Physics and Astronomy, University of Edinburgh, Edinburgh, EH9 3FD, United Kingdom}
         \affiliation{Departamento de F\'isica, Escuela Polit\'ecnica Nacional, Quito, 170143, Ecuador}

        \author{Liang Chen}
        \email{bqipd@pm.me}
        \affiliation{School of Fundamental Physics and Mathematical Sciences,
Hangzhou Institute for Advanced Study, UCAS, Hangzhou 310024 ,China}
        \affiliation{University of Chinese Academy of Sciences, 100190 Beijing, China}
        \author{Thomas W. Kephart}
        \email{tom.kephart@gmail.com}
        \affiliation{Department of Physics and Astronomy, Vanderbilt University, Nashville, TN 37235, USA}

\begin{abstract}
Stimulated emission is shown to be robust
in stars.  Through Bose enhancement this produces quantum
states of aligned, monochromatic photons somewhat similar to
a laser.  The probability of creating such
states is computed.
We show that from the solar corona
such quantum states would propagate outside of the solar
region and through the Solar System without decoherence.
For a $1 {\rm m}^2$ test detector at the distance of the Earth from the Sun, we
estimate rates of such quantum states
in the few per second thus potentially detectable.
The same process
should lead to such quantum states also arriving from stars at 
interstellar distances.

\end{abstract}

\maketitle

\section{Introduction}

Recently it was observed by one of us \cite{Berera:2020rpl} 
that photons in certain
frequency bands can traverse interstellar
distances, without their initial quantum state decohering.
This fundamental point has potential far reaching consequences
for astronomical observation. That paper already noted   
applications both for interstellar quantum communication,
with further work along that line done in \cite{Berera:2022nzs}, and
possible detection of quantum coherent signals from
astrophysical sources. By quantum coherence of photons we mean very simply N-photons
in some specified quantum state, which in this paper in particular means
they are all identical,
so with the same momemtum $\vec{k}$ and polarization $s$. 
This, which we will call a N-identical photons state, is
suggested as a possible type of quantum coherent
photonic signal.  In this paper we identify a specific   
mechanism in stars and their atmosphere that
can create such states and examine the possibility for observing these quantum coherent photon signals at distances far
away from emission.

The elementary process creating a N-identical
photons state is stimulated emission.
When a photon in state $\ket{1_{\vec{k},s}}$ impinges upon an atom that is in an excited state
of the same energy as itself, an emission from the
atom is possible that then leads to two photons in the same state, $\ket{2_{\vec{k},s}}$.
This two–photon quantum state will be identified as a two–identical photons state.
The process can repeat by impinging these photons on another atom
to create a three-identical photons state with a Bose enhancing
factor and so forth.
We observe that in the stellar environment there should be a ubiquitous
production of such N-identical photons states, especially the simplest
two photon process.

Deep enough below the surface of the star, these
states will quickly decohere through interactions. However
if such a stimulated emission event occurred above the
surface in the stellar atmosphere, then it is possible these photons could 
escape with their quantum state intact.

In this paper we will examine the details of such stimulated emission
events inside the stellar environment. We will then examine the
mean free path (MFP) of photons both in the stellar environment
as well as in the interplanetary region of the Solar System
and assess the flux of such photonic quantum states that could
be measured.  We will then discuss ways in which such
states could be detected both directly and indirectly.

Here we identify all potential sources of decoherence and show that these effects are small. We will take a first
principles approach and look directly at the interactions of the
photons with all particles in the medium and show the mean
free paths from such interactions are much longer than the distances the photons are required to traverse, often orders
of magnitudes longer.  Of course there may be some residual
decoherence and before attempting experiments, this would need to
be understood. Standard methods such as density
matrix,
master equation etc., could be employed to quantify these higher order
effects. The purpose of this first paper is
to highlight the dominant physical effects and establish that to
zeroth order decohering effects are negligible.

\section{Mechanism of stimulated emission}

Let us start with a brief review of the mechanism of stimulated emission.
A more detailed discussion of it can be found in textbooks 
such as \cite{townsend_2012}.  
To calculate the transition probability of stimulated photon emission, 
using Fermi's Golden Rule gives the matrix element 
$
\bra{(N+1)_{\vec{k},s}} \otimes \bra{\Psi_f} H_I \ket{\Psi_i} \otimes    \ket{N_{\vec{k},s}}
$,
where $\ket{\Psi_i}$ and $\ket{\Psi_f}$ are initial and 
final states of the atom, respectively, 
$\ket{N_{\vec{k},s}}$ is the $N$-photon state in Fock space 
with momentum $\vec{k}$ and polarization $s$, and $H_I$
is the interaction Hamiltonian between the electromagnetic field and atom. 
Such a matrix element will lead to a
quantum mechanical Bose enhancement factor such as
$
\bra{(N+1)_{\vec{k},s}}  \hat{a}^\dagger_{\vec{k},s}   \ket{N_{\vec{k},s}}  = \sqrt{N+1}  ~,
$
which means that for a state with a high number of photons, there is an
enhanced chance of stimulated emission of one more photon 
in the same state.

A semiclassical approach such as the one in \cite{svelto_2010} leads to an expression for the cross 
section of the stimulated emission off a single atom of the form
\begin{equation}    \label{cross-section-se}
\sigma  =   \frac{  2\pi^2} { 3 n \epsilon_0  c  h } |\mu|^2  \nu  g(\nu - \nu_0)  ~.
\end{equation}
Here, $n$ is the refractive index of the medium, $\mu$ is the magnitude of 
the time-independent electric dipole moment between the initial and final 
wave functions of the atom, $\nu$ is the frequency of the 
photon, and  $g(\nu - \nu_0)$ is the line shape function with 
width $ \Delta\nu_0$.  
In our discussion, we will choose the Lorentzian line shape function, 
though note that the results we find are comparable to 
those using other line shape functions such as Gaussian.

The stimulated emission process needs the electric dipole moment,
which can be extracted from the measured spontaneous emission rate.  
A textbook expression for the spontaneous emission rate reads \cite{svelto_2010}
\begin{equation}  
A  = \frac  {    16 \pi^3 \nu_0^3 n   |\mu|^2  }{   3   \epsilon_0  c^3  h  }  ~,
\end{equation}
with experimental values readily available on the NIST Atomic Spectra Database \cite{NIST}. 
The dipole moment $\mu$ appears both in the cross 
section \eqref{cross-section-se} of stimulated emission and in the
rate of spontaneous emission. Thus, we can express it in terms
of the measured rates $A$ without an explicit calculation.
This allows us to write Eq. (\ref{cross-section-se}) now as
$
\sigma   =  \frac   {  \lambda^2} {8\pi  n^2 }   g(\nu - \nu_0) A .
$
Substituting the peak value at $\nu_0$ of the Lorentzian line shape function, 
$
g(\nu_0 - \nu_0)= 2/(\pi \Delta\nu)\,,
$
into the expression for the cross section leads to the stimulated emission cross section at wavelength $\lambda_0$ of
$
\sigma_0    =    (\frac{\lambda_0}{2\pi n})^2 {A\over\Delta\nu}  .
$
As can be seen, the stimulated emission cross section depends on four quantities: the
wavelength $\lambda_0$, refraction index $n$ of the material, frequency 
width of the emission $\Delta\nu$, and rate of spontaneous emission $A$. 
Using the NIST data for the  numerical values of $A$ 
allows us to calculate 
cross sections of any stimulated emission line from any ions. Reexpressing the frequency width  $\Delta\nu$ to wavelength 
width $\Delta\lambda$,  we have
\begin{equation}  
\sigma_0     =   ({\lambda_0\over2\pi n})^2   ( { A \lambda_0 \over c } ) (  { \lambda_0 \over\Delta\lambda}) ~.
\end{equation}
Here $A \lambda_0/ c$ gives the ratio between the wavelength of the photon and 
the length the photon can travel during the time scale of spontaneous 
emission.  Moreover $\lambda_0 /\Delta\lambda$ gives the ratio 
between the wavelength of the photon and the line width.

\section{N-identical photons state probability from stimulated emission}

Let's assume that we have an initial one-photon state $\ket 1$ at one side of a layer of medium of thickness $L$, filled with atoms of type $a$, with number density $n_a$. This initial single photon state $\ket 1$ travels through the layer from $x=0$ to $x=L$. By using coordinate label $x$, we are not trying to localize the photon but only want to find the probability of a $\ket 2$ state, a $\ket 3$ state or in general, an $\ket N$ state created through stimulated emission, if any measurement is performed at $x$.

Denote by $P(x, N)$  the probability of finding an $\ket N$ state at distance $x$, then the probability of finding an $\ket {N+1}$ state at distance $x+dx$ becomes $P(x+dx, N+1)$. The relation between $P(x+dx, N+1)$ and $P(x, N)$ is
\begin{flalign}  \label{prob}
& P(x+dx, N+1)
\\\nonumber
=& P(x, N+1) \left[ 1 -   (N+1) n_a\sigma_0 ~ dx \right]
+ P(x, N) (  N  n_a \sigma_0 ~ dx )~.
\end{flalign}
On infinitesimal scale $dx$, finding $\ket {N+1}$ at $x+dx$ includes two mutually exclusive events.
The first event is finding $\ket {N+1}$ at $x$ which carries probability $P(x, N+1)$ and then the $\ket {N+1}$ state does not induce stimulated emission from $x$ to $x+dx$, which carries probability $1 - n_a\sigma_0 (N+1) dx$. As we discussed above, the probability of a single stimulated emission event happening after a single photon traveling the distance $dx$ is $ n_a \sigma_0  dx$. Both quantum and semi-classical theory indicate that the probability of $\ket N$ induces a stimulated emission that is enhanced by a factor of $N$ compared to that of single photon state $\ket 1$. Thus, the probability of an $\ket {N+1}$ state inducing a stimulated emission event from $x$ to $x+dx$ becomes $ ( N+1 ) n_a \sigma_0  dx$, which needs to be subtracted to give the probability of not inducing stimulated emission.
The second event is finding $\ket N$ at $x$ which carries probability $P(x, N)$ and then the $\ket N$ state does induce stimulated emission from $x$ to $x+dx$, which carries probability $N  n_a\sigma_0 dx$. 

After solving equation \eqref{prob}, the probability of finding an $\ket N$ state at the other side of the layer is given by
\begin{equation}   \label{N-probability}
P(L, N) = e^{-n_a\sigma_0 L} \left( 1 - e^{-n_a\sigma_0 L} \right)^{N-1}\;.
\end{equation}
\noindent
As a sanity check, one can sum $P(L,N)$ over all $N$ from $1$ to $\infty$ to show that this expression is normalized to $1$, 
independent of $L$.

\section{Rate estimation}

Next, we estimate the rate of N-identical photons states being created at the solar corona, to then determine the number of such states that could be detectable, ideally in space,
on a $1\ {\rm m}^2$ area near the Earth. We shall consider photons in the optical, UV and X-ray regions.

For the optical, we obtain the rate of $\ket 1$ states from experimental data from \cite{165_Rusin}. Then, we calculate the MFP of stimulated emission, to finally compute $P(L,2)$ using the above formula. For example, Ref. \cite{165_Rusin} tracks the history of the intensity of the 530.3 nm line, which goes through yearly cycles. However, going to the lower end of the reported intensities, we can still have around  $2.4\times10^{12}$ photons arriving on an area of 1 m$^2$ per solid angle per second. 
(We conservatively picked a solar cycle of low corona index number plotted in \cite{165_Rusin} and converted the light intensity to photon number rate.)
Hence, $N_1^{530.3}\sim2.4\times10^{12}$ would conservatively count the rate of the $\ket 1$ state.

Let us now determine the number of $\ket 2$ states created from stimulated emission.
The particle density of the corona is at its lowest at the corona holes, therefore we conservatively take the particle number density $10^{9}$ m$^{-3}$ \cite{Morgan_2020} of the corona holes as the density of the entire corona. Schmelz \cite{Schmelz_2012} et al. gives an abundance of iron in the corona of $7.08\times10^{-5}$ with respect to hydrogen. This yields an iron number density of $7.08\times10^{4}$  m$^{-3}$. The ionization fraction  of Fe XIV-530.3 nm depends on temperature and peaks at around 0.2 for a temperature $T=2\times 10^6$ K  based on \cite{Habbal_2011}. Conservatively, we choose the density of Fe XIV at the excited level of 530.3 nm to be $1.0\times10^{4}$ m$^{-3}$.

Given the cross section of stimulated emission
$
7.585 \times 10^{-25} \text{ m}^2~,
$
an upper limit of the MFP of stimulated emission of 530.3 nm photons in the corona is
$
\ell_{mfp}^{530.3} =1.32\times10^{20}\text{ m}~.
$
Then, if, based on above, a number $N_1^{530.3}=2.4\times10^{12}$   of 530.3 nm $\ket 1$ states are emitted on the 1 m$^2$ area per solid angle every second, the number of $\ket 2$ states emitted on the same area per solid angle every second can be computed through \eqref{N-probability} to be
$
N_2^{530.3}  
 \sim ~ 145.6
~.
$

Similar calculations show that states equal or higher than $\ket 3$ are negligibly produced.
Although we have conservatively estimated the following quantities including
the number of $\ket 1$ states of $530.3$ nm, the overall corona density, the
relative abundance of iron in the corona, and the fraction of Fe XIV ions of the
specific energy level, there are over 100 $\ket 2$ states per second per
solid angle emitted onto the 1 m$^2$ area near the Earth.

Turning now to the UV and X-ray emission,  we will first estimate the rate of $
\ket 1$ states. Given the available experimental data, we shall use a different approach to compute the densities of the excited ions that will lead to the photons at those energies. For this, the following formula for the intensity of a line will come in handy,
 \begin{equation}\label{intensity}
    I_{\rm line} = \int_0^{\infty} I(\lambda) d\lambda = \frac{1}{4\pi} \frac{hc}{\lambda_0} A N_k L\;,
\end{equation}
where $A$ is the atomic transition probability we encountered previously, and $N_k$ is the number per unit volume (number density) of excited atoms in the upper (initial) level $k$. Based on the measurements in \cite{coronathick}, the thickness of the corona is about $8$ million kilometers, which we will take as the value of  $L$. Ref. \cite{Young_2008} gives photon intensities of several lines from Fe XII and Fe XIII in the UV range obtained via satellite instruments. With the formula above, we seek to estimate the rate of $\ket 1$ states from measurements made in \cite{Young_2008} and \cite{1985ApJ...297..338R}, for UV and X-ray respectively.

The intensity data of 19.664 nm line of Fe XII in \cite{Young_2008} is over 800  erg$/$(cm$^2$ s \AA) at the peak. We pick a value of 600, (since there is a 19.654 nm peak of 600 and we are unsure of any potential overlap between the two) multiply it with a wavelength width, convert the unit to eV$/$(m$^2$ s), and then divide by the energy of 19.664 nm photon in eV. This indicates that $N_1^{19.664}\sim5.9\times10^{14}$ of $\ket 1$ states arrive on the area per second. The flux data in \cite{1985ApJ...297..338R} shows that the rate can be $N_1^{1.5}\sim3.7\times10^{10}$ for $\ket 1$ states of the 1.5 nm line of Fe XVII  on the same area.

By reversing \eqref{intensity} and using these intensity and flux data aforementioned \cite{Young_2008, 1985ApJ...297..338R}, we find the density of Fe XII and Fe XVII that are excited to the 19.664 nm and 1.5 nm level, are around $1.9\times10^{-5}$ m$^{-3}$ and $2.6\times10^{-12}$ m$^{-3}$, respectively.
Following the same approach by which we estimated the rate of $\ket 2$ for 530.3 nm, using $P(L,2)$, we find the rate of $\ket 2$ for 19.664 nm and 1.5 nm to be around $N_2^{19.664}\sim2.83$ and $N_2^{1.5}\sim 0$ on the same area.  As the frequency of the photon increases, less $\ket 2$ states can be created from this stimulated emission process, which is not surprising since the cross section, Eq (3), goes as  $\nu_0^{-4}$.

There are many uncertainties in the understanding of the solar corona,
which have allowed us to only make rough estimates of the rates
of quantum states from this stimulated emission process.  Nevertheless, we think that the key point is we are finding rates
at a few per second in our test detector, rather than on much longer
time scales, thus in the realm of being measurable.

\section{Decoherence}

Having discussed a mechanism to produce this kind of N-identical photons state, one has to wonder how likely it is for it to get to our detector intact without decoherence. As it turns out, the possibility of decohering interactions, in particular for the extreme ultraviolet to the X-ray range, is basically null. In spite of the peculiarities of the corona, current observations do allow us to put strong bounds on MFPs and interaction rates of the potentially dangerous decohering interactions. 
Spectroscopic studies have shown that there are three distinct ``regions'' in the corona, dubbed as K, F and E. The K-corona shows no absorption, so there is a continuum of white light that undergoes Thomson scattering, which could affect our quantum coherent photonic state. The optical depth for that interaction is $\tau = \int dr \sigma_{\rm Th} n_e$, where $\sigma_{\rm Th} = 6.65 \times 10^{-29}\ {\rm m}^2$ is the Thomson cross section and $n_e$ is the electron density in the corona, for which there are several models. Those by Allen and Baumbach \cite{wexler2019spacecraft}, and Edenhofer, et al. \cite{edenhofer1977time} respectively state 
\begin{eqnarray}
    n_e(r) & = &\left[\frac{2.99}{r^{16}} + \frac{1.55}{r^6} \right] 10^{14}\ {\rm m}^{-3}\;, \quad 1.2 \lesssim r < 3\;, \label{eq:ne1}\\
    n_e (r) &=& \left[\frac{30}{r^6} + \frac{1}{r^{2.2}} \right] 10^{12}\ {\rm m}^{-3}\;, \quad  3 < r < 65\;, \label{eq:ne2}
\end{eqnarray}
where $r$ is in units of $R_{\odot}$. With that information, we arrive to $\tau \simeq 5 \times 10^{-7}$, which corresponds to an interaction probability of $6 \times 10^{-5}\%$, and a MFP of $10^{17}$ m, almost 6 orders of magnitude longer than the distance between the Earth and the Sun. 
Notice that other models will lead to the same order-of-magnitude densities, so our conclusions would be the same \cite{wexler2019spacecraft}. 

Next, we shall consider potential interactions with the elements in the corona. Naturally, there is less information about the distribution of the atoms throughout the corona, but we can still (over)estimate the probability of each interaction using available data. As previously mentioned, we know the abundances with respect to hydrogen. Unlike before, let us now fix the particle density of the corona to be about $10^{15}\ {\rm m}^{-3}$, which we assign to hydrogen. In doing so, the density of each species is overinflated as well as the interaction probabilities.
These overestimates are done
in order to give us the shortest estimates for the MFPs from decohering processes. Using the abundances reported in \cite{Schmelz_2012} and total attenuation cross sections (coherent and incoherent scattering + photoelectric absorption + \ldots) from the XCOM software from \cite{NIST}, we obtained a lower value estimation of the MFP due to photon-hydrogen interactions of $\ell = (\sigma n)^{-1} \simeq 8 \times 10^{11}$ m, which is 5 times larger than the distance between the Sun and Earth. Notice that we chose photon energies $\sim 1$ keV, where the cross section always peaks. MFPs of the same order of magnitude are obtained for interactions in the corona with He ($4 \times 10^{11}$ m), Mg ($6 \times 10^{11}$ m), Si ($9 \times 10^{11}$ m), C ($10^{12}$ m), N ($2\times 10^{12}$ m), Ni ($3\times 10^{12}$ m), or S ($5 \times 10^{12}$ m). The smallest MFP is obtained for O ($2 \times 10^{11}$ m), whereas for other elements they are larger than those shown above, and among those the strongest interactions are with Al, Cr and Mn, which in turn render $\ell \simeq 10^{13}$ m. 
All these MFPs must be combined, with an effective or total MFP given by 
\begin{equation}
    \frac{1}{\ell_{\rm eff}} = \sum_i \sigma_i n_i = \sum_i \frac{1}{\ell_i}\,,
\end{equation}
where the sum is over all the atoms under consideration. Upon performing this summation, we find:
\begin{eqnarray}
    \ell_{\rm eff} = 4.6 \times 10^{10}\ {\rm m}\,.
\end{eqnarray}
This value lies within an ${\cal O}(1)$ factor of the distance between the Sun and the Earth, and, most importantly, it exceeds the width of the corona by at least 2 orders of magnitude. This reaffirms the validity of our previous conclusions that decohering effects are
negligible.

Another decoherence factor could be the dust near the corona, which plays a crucial role in the F region. The process of interest is the scattering of photons off dust particles, which largely depends on their size. Fragments smaller than $1\ \mu$m are not expected near the corona since they are pushed away by radiation pressure and electromagnetic forces \cite{mann2019dust}. For this reason, the brightness of the F-corona, mostly of a thermal nature, is dominated by the near–IR range of the electromagnetic spectrum \cite{kimura1998brightness}. For radiation of shorter wavelengths, one can compute $\ell = (\sigma_{\rm geom} n)^{-1}$, where $\sigma_{\rm geom}$ is the geometric cross section of the dust particles. The average flux detected by Helios \cite{kruger2019interstellar} in a range of heliocentric distance from $0.3$ to $1.0$ AU was of $(2.6 \pm 0.3) \times 10^{-6}\ {\rm m}^{-2}\ {\rm s}^{-1}$, with dust particles of size $0.37\ \mu$m. Taking the worst case scenario, the flux at the photosphere is
\begin{equation}
    \Phi = 2.6 \times 10^{-6} \left(\frac{1\ {\rm AU}}{R_{\odot}}\right)^2 {\rm m}^{-2}\ {\rm s}^{-1} \approx 0.12\ {\rm m}^{-2}\ {\rm s}^{-1}\;,
\end{equation}
corresponding to an interaction rate $\Gamma = \Phi \sigma_{\rm geom} \simeq  5.16 \times 10^{-14}\ {\rm s}^{-1}$ and a MFP of $6 \times 10^{21}$ m. Similarly, the Solar Orbiter estimated a flux of $8 \times 10^{-5}\ {\rm m}^{-2}\ {\rm s}^{-1}$ at 1 AU for particles of size of order $0.1\ \mu$m \cite{zaslavsky2021first}. A similar calculation  leads to an interaction rate of $10^{-13}\ {\rm s}^{-1}$, and a MFP of order $10^{21}$ m, the same as our previous estimate. 

As large as these MFPs are, they are relatively short in comparison to those associated to interactions with the blackbody radiation from the photosphere, where $\ell \sim 10^{43}\ {\rm m}$. This is due to the weakness of the interaction, with the cross section dependence as $\alpha^4 \sim 137^{-4}$. 

One potential decohering effect that can not be assessed through direct particle interactions
is Faraday rotation.
To evaluate the impact of the Faraday effect, we can estimate the rotation measure angle $\beta$, which quantifies the effect for a given wavelength $\lambda$. It is given by the expression 
$$\beta = \frac{e^3}{8\pi^2 \epsilon_0 m^2 c^3} \lambda^2 \int_{R_{\odot}}^d n_e (s) B_{\parallel} (s) \mathrm{d}s\,,$$
where $n_e(s)$ represents the electron density in the medium and $B_{\parallel}(s)$ denotes the component of the magnetic field along the trajectory \cite{ferriere2021correct}. For an upper bound approximation, we can utilize equations \eqref{eq:ne1} and \eqref{eq:ne2}, together with the assumption of $B_{\parallel} \sim 100\ \mu$T, which is an exaggerated approximation for the entire spatial range under consideration \cite{kooi2022modern}. This simplified calculation yields
$$\beta(\lambda=100\ {\rm nm}) \simeq 10^{-18}\,.$$
Hence, the overall effect of Faraday rotation on the state is exceedingly small and would impact every photon uniformly. This is due to the expectation that the photons are part of the same wave train, with a microscopic separation, whereas the coherence lengths of the magnetic fields involved are macroscopic. Consequently, this effect leads to an overall rotation without any decoherence effects. 

Another effect, birefringence, though typically addressed in terms of the wave properties of photons, ultimately still arises from the collective interactions of the photons with the particles in the medium.
The particle analysis we conducted already accounts for these interactions, rendering long MFPs. 
While our analysis employed the Thomson scattering formula instead of the more intricate Klein-Nishina expression, our conclusions are highly robust, because the Thomson cross section is always larger than the Klein-Nishina one. As a result, we can confidently extrapolate our findings to higher dispersion energies. This assertion applies equally to the Faraday effect, as far as decoherence goes. Thus, the lack of interactions from a particle perspective results in an effective transparency from a wave point of view. Finally, it is important to note that the occurrence of the Faraday effect does not necessarily imply decoherence of the state. Similar to our study on effects from
gravity \cite{Berera:2022nzs}, it suggests a change in fidelity, which should be distinguished from the loss of quantum coherence, a crucial differentiation for our specific objectives.

Other potential decoherence factors have been discussed elsewhere \cite{Berera:2020rpl,Berera:2021xqa, Berera:2022nzs}. There, it was found that photons at the energy ranges discussed here would not interact with the cosmological medium, or the cosmic background of photons in the microwave, IR, and X-ray bands, with other interactions depending on the particular local environments the photon travels through. For example, X-ray photons could travel through the interstellar medium without interacting with other particles for distances up to 1 Mpc depending on the region, although this can be reduced to 100 pc for dense HII regions. Factors like the matter environment of the Solar System have also been considered, yielding almost null probabilities of interactions. 

Finally, it is worth noting that the quantum nature of a photon state is also unaffected by gravity (see \cite{Berera:2022nzs} and references therein). The effects of gravity on quantum states have been extensively explored in the literature, especially in the context of quantum communications and the potential applications stemming from them. Some examples include the effects of gravity on quantum information protocols \cite{Bruschi:2013sua}, frequency spectrum deformation \cite{Bruschi:2021all}, gravitational distortion of quantum communication \cite{Kohlrus:2015szb,Exirifard:2020yuu}, geometric phase acquired as a wave-packet travels over a null geodesic \cite{9597920,Exirifard:2021sfc}, and gravitational redshift induced transformations on the photon state \cite{Bruschi:2021rhk, Rodriguez:2023par}.
In the context of this work, it is crucial to recognize that while the photon state may experience changes affecting features like the fidelity in quantum communications, its fundamental quantum nature remains unaltered. As a result, in the case of the N-identical photons states discussed in this paper, any phase effects induced by gravity will be uniform among all photons within that state, ensuring the preservation of their quantum integrity up to an overall phase.
Consequently, even though our analysis primarily focuses on the solar corona, it is plausible that signals originating from other stars can also reach our detectors near Earth essentially intact, especially in the UV and X-ray range.

\section{Measurements}

In principle, measuring the quantum nature of a signal from space could be
significantly more challenging than in a controlled setup. However, one
could potentially use the same principles that are common in Earth-based
experiments. For one, direct measurements of the photons could be made through interference experiments that induce a phase
between split signals leading to characteristic correlations depending on their
nature. 
For example, the Hong-Ou-Mandel (HOM) effect \cite{Hong:1987zz} could be used to test
the indistinguishability of the two photons. The effect occurs when
two identical single photons enter a 1:1 beam splitter through different
input gates, and under certain temporal conditions, the two photons will
always exit the beam splitter together through the same terminal.
This effect has been used in an astrophysical experiment by
Deng et al. \cite{Deng:2019der,duanadded} for testing quantum interference between
single optical photons from the Sun and a quantum dot on the Earth,
to provide distinct evidence for
the quantum nature of thermal light. 
Other astrophysical quantum experiments in the optical region have
also been proposed by Dravins et al. \cite{2008ASSL..351...95D,dravins2005quanteye}.
Our above analysis has shown to minimize environmental decoherence, the UV and X-ray are
good regions for testing our proposed mechanism.
Alternatively, indirect measurements could be made through examining the distribution probability of thermal signals, which would defer
from that of quantum signals. For example, the states emitted by a laser follow
 Poisson statistics, where the probability of measuring $N$ photons in a
given coherent state peaks significantly around the mean value. On the
contrary, for thermal sources the largest probability always corresponds to
the no-photon state, with the distribution function decreasing less rapidly
than for coherent states. Thus, provided that detection at the photon level
can be performed, current technology is well suited to discern a quantum
signal from a classical one, but extending into the UV and X-ray bands and to our particular type of N-identical photons states still needs further development.

\section{Conclusion}

This work has identified the emission of specific types of quantum
states of photons, N-identical photons states, originating from specific processes,
stimulated emission, in the
atmosphere of stars.  Focusing on the Sun, we showed that arriving at a $1 {\rm m}^2$ detecting region near the Earth the rate of such N-identical photons
states would be a few per second thus should be measurable.
An actual experiment could of course be done at a distance closer to the Sun, which would increase signal flux and reduce
residual decoherence and some possible experimental errors.
We also showed such states would not undergo decoherence both leaving
the solar environment and propagating from the Sun to Earth, especially
in the UV and X-ray regions.
From our previous
work \cite{Berera:2020rpl,Berera:2021xqa,Berera:2022nzs}, 
we expect that such states can also
arrive without decoherence from stars at interstellar distances away.

It is remarkable how the creation of such quantum states can be
specifically identified within an astrophysical body and
that such states then can traverse astronomical distances without
decohereing.
In fundamental terms this work has extended the distance of measurable quantum phenomenon, and that at the individual particle level, to astronomical scales.
In practical terms these results provide a new probe
for studying stellar atmospheres.

\section*{Acknowledgments}

AB is partially funded by STFC. JCF thanks Nicol\'as V\'asquez and Christian V\'asconez for helpful discussions about dust in the corona. This work is supported in part by the National Key Research and Development Program of China under Grant No. 2020YFC2201501 and  the National Natural Science Foundation of China (NSFC) under Grant No. 12147103.

\bibliography{apssamp2}

\end{document}